\documentclass[]{aa}

\usepackage{graphics}
\usepackage{txfonts}

\begin{document}

\title{Non-LTE Balmer line formation in late-type spectra:  \\ Effects of atomic processes involving hydrogen atoms  
} 
\author{P. S. Barklem
}
 
\offprints{P. S. Barklem,
\email{barklem@astro.uu.se}}

\institute{Department of Astronomy and Space Physics, Uppsala University, Box 515, S 751-20 Uppsala, Sweden 
}

\date{Received 3 November 2006 / Accepted February 2007}

\abstract
{The wings of Balmer lines are often used as effective temperature diagnostics for late-type (F, G, K) stars under the assumption they form in local thermodynamic equilibrium (LTE).  
}
{Our goal is to investigate the non-LTE excitation and ionisation of hydrogen and the formation of Balmer lines in late-type stellar atmospheres, to establish if the assumption of LTE is justified.  Furthermore, we aim to determine which collision processes are important for the problem; in particular, the role of collision processes with hydrogen atoms is investigated.}
{A model hydrogen atom for non-LTE calculations has been constructed accounting for various collision processes using the best available data from the literature.  The processes included are inelastic collisions with electrons and hydrogen atoms, mutual neutralisation and Penning ionisation.  Non-LTE calculations are performed using the \texttt{MULTI} code and the MACKKL semi-empirical solar model, and the relative importance of the collision processes is investigated.  Similar calculations are performed for \texttt{MARCS} theoretical models of other late-type stellar atmospheres.}
{Our calculations show electron collisions alone are not sufficient to establish LTE for the formation of Balmer line wings. Mutual neutralisation and Penning ionisation are found to be unimportant.  The role of inelastic collisions with neutral hydrogen is unclear.  The available data for these processes is of questionable quality, and different prescriptions for the rate coefficents give significantly different results for the Balmer line wings. }
{Improved calculations or experimental data are needed for excitation and, particularly, ionisation of hydrogen atoms in low-lying states by hydrogen atom impact at near threshold energies.  Until such data are available, the assumption of LTE for the formation of Balmer line wings 
in late-type stars is questionable.}


\maketitle

\section{Introduction}
\label{sect:intro}

For more than 80 years it has been known that hydrogen is the dominant chemical element in the Sun, and most other stars, by number of atoms as well as by weight.  For the Sun, and even more so for more metal-poor stars of solar type, hydrogen is the dominant continuous opacity source through H$^-$ and \ion{H}{i} bound-free and free-free absorption, as well as Rayleigh scattering. For the metal-poor stars of solar temperatures, hydrogen is also important as an electron donor and thus governs the H$^-$ opacity.  In addition, the hydrogen lines in late-type stars are important as temperature diagnostics, and the Balmer discontinuity as a gravity diagnostic for F stars.

When the development of electronic computers and suitable algorithms in the 1960's made it possible to reasonably self-consistently model excitation and ionisation together with radiative fields in stellar atmospheres, these advances were immediately applied to the case of hydrogen.  Most studies considered early-type stars, but the situation for late-type stars was also explored (see Mihalas \& Athay~\cite{mihalas73} for a review of the early development). The newly developed Feautrier method was used by Cuny~(\cite{cuny67}) to model hydrogen in the solar atmosphere with a 3--5 level model atom with continuum.  The H$\alpha$ line was found to be controlled by the photoionization processes from its lower and upper level, and to be sensitive to the structure of the photosphere. Strom~(\cite{strom67}) studied the effects of departures from local thermodynamic equilibrium (LTE) for hydrogen on continua of G and K stars. The equations of statistical equilibrium were solved for the three lowest levels (assuming detailed balancing in the lines) and H$^-$.  Observable effects on fluxes and colours were found, but these vanished when more realistic values for the rate constant of the important associative detachment reaction, $\mathrm{H}^- + \mathrm{H} \rightleftharpoons \mathrm{H}_2 + e$, were used (Lambert \& Pagel~\cite{lambert68}; Pagel~\cite{pagel68}). 

Vernazza et~al.~(\cite{vernazza81}) calculated the statistical equilibrium of hydrogen in their semi-empirical solar models using a 12-level model atom with continuum.  They included resonance and van der Waals broadening, but neglected Stark broadening, as did Cuny~(\cite{cuny67}).  Departure coefficients\footnote{In this paper we define the departure coefficient of level $i$ as $b_i=n_i/n_i^*$, with $n_i$ the population and $n_i^*$ the LTE population.  Some other authors use the definition $b_i=(n_i/n_i^*) / (n_\kappa/n_\kappa^*)$, where $\kappa$ is the ionised state.} $b$ significantly different from unity were found in the upper photosphere.  When comparing their hydrogen line profiles to observations, they found rather severe discrepancies for Ly$\beta$; for H$\alpha$, the results in the core were shown to be dependent on the number of levels used as well as on the particular model atmosphere. Vernazza et~al.\ also included statistical equilibrium calculations for H$^-$ in their solar models.  They found departure coefficients that are significantly greater than unity for the continuum optical depths $\log \tau_{500\mathrm{nm}}< -5$, and 0.9 to 1.0 in the depth range $-4 < \log \tau_{500} < -2$, three times greater than those found by Lambert \& Pagel~(\cite{lambert68}).  The differences seem to be mainly caused by the fact that the temperature minimum of the solar model atmosphere used by Vernazza et~al.~(\cite{vernazza81}) (their model C for the average quiet Sun) is deeper and located further out than in the model used by Lambert \& Pagel~(\cite{lambert68}).  At $\log \tau_{500} > -2$ the $b_{\mathrm{H}^-}$ coefficients found by Vernazza et~al.~(\cite{vernazza81}) are unity, verifying the conclusion of Lambert \& Pagel~(\cite{lambert68}) that LTE is a good approximation for the calculation of H$^-$ continua.

Recent improvements in high-resolution stellar spectroscopy have made it possible to obtain observed stellar Balmer line profiles with reasonable accuracy, stimulating further study of these lines.  The formation of the Balmer line wings occurs at relatively large optical depths compared to most of the spectral lines in late-type stars.  This is due to the high excitation energy of the $n=2$ level, and makes it possible to probe the upper convection zone of the atmospheres.  Fuhrmann et~al.~(\cite{fuhrmann93}) compared observed Balmer line profiles for late-type dwarfs with calculated profiles, based on the LTE approximation.  They found discrepancies which were attributed to effects of inadequate temperature structures, as obtained with a mixing-length description of the convective energy transport, and suggested revisions of the mixing-length parameters.  In following papers (e.g.\ Fuhrmann~et~al.~\cite{fuhrmann94}), fitting of Balmer line wing profiles was used to obtain effective temperatures.  A number of studies of Balmer line profiles in late-type stars in relation to convective parameters and temperature scales followed; for example, van't Veer-Menneret~et~al.~(\cite{vantVeerMenneret98}), Gardiner~et~al.~(\cite{gardiner99}), and Barklem et~al.~(\cite{barklem02}).  

All these recent studies, have assumed that the Balmer wings form in LTE.  Vernazza et~al.\ calculated departure coefficients of unity for all levels below $\log \tau_{500} \sim -0.5$; however, departures are seen in higher regions of the atmosphere.  Similar results were found by Carlsson \& Rutten~(\cite{carlsson92}) in their study of solar Rydberg lines.  Both of these studies included only collisions with electrons in the model atoms, and did not compare computed Balmer lines with those computed in LTE.  As the Balmer line wings typically form in the region around $-2 \la \log \tau_{500} \la 0.5$,  based on these results one would expect departures from LTE in Balmer line wings.  The assumption that collisions with electrons are the dominant collision process is based on the fact that at the thermal velocities of interest, electron collisions are expected to be non-adiabatic and have substantial cross sections, while collisions with heavier particles, such as H, are expected to be nearly adiabatic and thus have small cross sections.  Such arguments can be made from the Massey criterion~(Massey \cite{massey49}, see also Anderson~\cite{anderson81} for an illustrative discussion), comparing the typical collision time with the inverse of the natural frequency of the transition (given by $h/\Delta E$, the period of light corresponding to the transition).  If the collision time is similar to this period then the collision is non-adiabatic and a transition is relatively likely.  If the collision time is much larger than this period then the collision is nearly adiabatic and a transition is expected to be unlikely.  Furthermore, the high thermal velocity of electrons leads to a higher collision rate.  However, even if the inelastic cross sections for collisions with heavy particles such as hydrogen atoms are indeed small, the abundance of perturbers is extremely large.  In the photosphere of solar-type stars neutral hydrogen atoms typically outnumber electrons by a factor of $10^4$, and can be orders of magnitude greater still in metal-poor stars.

A recent study of the solar Balmer and Paschen lines by Przybilla \& Butler~(\cite{przybilla04b}) did compare LTE and non-LTE line profiles and found them to be identical in the wings.   While their model atom employed improved collision rates for electrons, it also included collisions due to hydrogen following the semi-empirical formulae of Drawin (\cite{drawin68, drawin69}) empirically scaled by a factor of two in order to reproduce the observed solar line profiles. 

Thus, it has never been established that Balmer line wings in late-type stellar spectra are formed in LTE.  The basis of the assumption is that the line wings form deep in the atmosphere where collisional processes are assumed to be efficient and dominant over radiative processes.  Even if this turns out to be the case, the atomic processes responsible have never been identified.   In this work, we make a non-LTE study of hydrogen and the formation of Balmer lines in late-type stellar atmospheres.  In particular, we examine a number of collisional processes, especially those involving perturbing hydrogen atoms, and their effects on the statistical equilibrium of hydrogen using available estimates of the collisional transition probabilities.  Modern line opacity data are employed.  In order to identify the important collision processes and examine the validity of the LTE approximation in the calculation of Balmer lines for late-type stars, we have undertaken statistical equilibrium calculations in the semi-empirical solar model of Maltby~et~al.~(\cite{maltby86}) and \texttt{MARCS} theoretical model atmospheres of dwarfs and giants of different metallicity (\S~\ref{sect:stat_equil} and~\ref{sect:nonlte_results}).   
Finally, we discuss the results and our conclusions (\S~\ref{sect:discussion}).

\section{Statistical-equilibrium calculations}
\label{sect:stat_equil}
 
The equations of statistical equilibrium and radiative transfer are solved for a \ion{H}{i} model atom with 19 bound levels and continuum in a 1D plane-parallel model atmosphere, consistently with the hydrostatic equilibrium equation, using the \texttt{MULTI} code (Carlsson~\cite{carlsson86}, Carlsson~et~al.~\cite{carlsson92b}).    
The atomic data for the model atom will be described in detail below.

The calculations assume complete redistribution of the line radiation for hydrogen lines.  Hubeny \& Lites~(\cite{hubeny95}) studied the effects of partial redistribution on hydrogen in the Sun.  Their method was also applied by Sim~(\cite{sim01}) to other late-type stars, including Procyon (F5 IV-V), $\beta$ Gem (KO III) and $\alpha$ Tau (K5 III).   Partial redistribution is particularly important in strong resonance lines, such as Ly$\alpha$.   Although the direct effects on the line profile are most obvious, and lead to effects, for example, on the ionization balance, some more secondary effects also appear via the populations of excited states. Partial redistribution effects may also be significant via ``cross-redistribution'' (or resonant Raman scattering) between Ly$\beta$ and H$\alpha$, sharing the same upper level.  That is, a photon may scatter in one line and reappear in the other before the atom undergoes a redistributing collision. These effects were included in the calculations of Hubeny \& Lites, and in those of Sim, but they do not include partial redistribution directly in the subordinate lines like H$\alpha$, arguing that the effects of partial redistribution in these lines are negligible for the conditions in the solar atmosphere (see also Hubeny \& Heinzel~\cite{hubeny84}).

\subsection{Model atom}

\subsubsection{Energy levels and radiative transition data}

We employ a 19-level-plus-continuum \ion{H}{i} model atom based on the model atom of Carlsson \& Rutten~(\cite{carlsson92}), from which all energy levels and radiative transition probabilities are adopted. Numerical experiments suggest that results converge with increasing number of levels (e.g. Przybilla \& Butler~\cite{przybilla04a}), and for solar-type stars 19 levels is found to be more than sufficient to achieve such convergence in line forming regions. 

Detailed line profiles are required for the statistical equilibrium and radiative transfer calculations.  The hydrogen line profiles are calculated including Stark broadening, self-broadening, fine structure, radiative broadening, and Doppler broadening (both thermal and turbulent).  The line profiles of the Balmer lines of particular interest here (H$\alpha$, H$\beta$ and H$\gamma$) are described as accurately as possible, including an accurate convolution of the profiles resulting from the different broadening mechanisms.  The Stark broadening profiles from Stehl\'e \& Hutcheon~(\cite{stehle99}) are employed, but if the plasma parameters ($T$, $N_e$) exceed the range for which Stehl\'e \& Hutcheon have tabulated line profiles, the Stark broadening profiles from Vidal et~al.~(\cite{vidal73}) are used instead.  Self-broadening from Barklem et~al.~(\cite{barklem00b}) is used.    The Gaussian (Doppler) and Lorentzian (self-broadening and radiative) components are convolved together employing the Voigt function, which is then convolved with the fine structure.  The resulting profile is then convolved with the Stark broadened profile using the Fast Fourier Transform technique for computational speed.

All other hydrogen line profiles are calculated in a more approximate manner for fast computation.  We use a slightly revised version of the code \texttt{HLINOP} which has been described by Barklem \& Piskunov~(\cite{barklem03}), which is based on the original \texttt{HLINOP} by Peterson \& Kurucz \footnote{see http://kurucz.harvard.edu/}.  In this case, the Stark broadening is calculated using the theory of Griem (\cite{griem60} and subsequent papers) with corrections based on Vidal et~al.~(\cite{vidal73}).  Self-broadening is included following the resonance broadening theory of Ali \& Griem~(\cite{ali66}).  The profiles are convolved in an approximate manner.  In the line core the Stark profile is approximated by a Lorentzian, and thus convolution is approximated via a Voigt profile.  In the line wings, profiles are simply added together.  We made test calculations comparing the results for H$\alpha$ and H$\beta$ if computed in the detailed manner described above, with the results where the lines are computed with the more approximate method just described.  The differences in the computed line wings for solar-type stars are very small, less than 0.08 \% and 0.3 \% in relative flux for H$\alpha$ and H$\beta$, respectively, in the solar case.  The differences in the line cores are more substantial, particularly in the widths of the cores; at the sides of the cores the differences are of the order of 5 \% and 2 \% for H$\alpha$ and H$\beta$, respectively.  Thus, a more careful convolution of profiles is needed for accurate calculation of the line cores.

\subsubsection{Collisional transition data}

The collisional transition data has been significantly updated and expanded from the model atom of Carlsson \& Rutten~(\cite{carlsson92}), which considered only collisions with electrons.  We added additional collisional processes in order to assess their importance.  These processes and their modelling are detailed below. 

For some processes, rate coefficients could be taken directly from the literature, and in these cases the references are provided and may be consulted for the relevant data.  However, in other cases we have derived rate coefficients from cross sections available in the literature via numerical integration over a Maxwellian distribution of velocities.  Where appropriate,  we provide the rate coefficients via a fit to numerical results of the form
\begin{equation}
\log \langle \sigma v \rangle = a_1 + a_2 \log T + a_3 (\log T)^2 + a_4 (\log T)^3. 
\label{eqn:ratefit}
\end{equation}
The fits are valid over the fitted temperature range of $1000 < T < 20000$~K.  

The collisional processes accounted for are now detailed.  Rate coefficients for reverse processes are calculated from the principle of detailed balance.  Throughout, $n$ denotes the principal quantum number of the level.  

\paragraph{Collisional excitation and ionisation by electrons}
\begin{eqnarray}
\mathrm{H}(n) + \mathrm{e} & \rightleftharpoons & \mathrm{H}(n^\prime) + \mathrm{e}
\end{eqnarray}
\begin{eqnarray}
\mathrm{H}(n) + \mathrm{e} & \rightleftharpoons & \mathrm{H}^+ + 2\mathrm{e}
\end{eqnarray}
Data from Przybilla \&\ Butler~(\cite{przybilla04a}), computed using the $R$-matrix method in the close coupling approximation, were used for transitions between states with $n\le 7$. For collisional ionisation, experimental results are available for ionisation from the 1s state from Shah~et~al.~(\cite{shah87}), and from the 2s state from Defrance~et~al.~(\cite{defrance81}), which are in good agreement with calculations, e.g., Bray \&\ Stelbovics~(\cite{bray93}) and Mukherjee~et~al.~(\cite{mukherjee89}), respectively.  We computed rate coefficients from the experimental cross sections; polynomial fits to the results are provided in Table~\ref{tab:ratefits}.  Calculations (see, e.g., Witthoeft~et~al.~\cite{witthoeft04}) indicate that the cross sections for excitation from the 2p state are of the same form as for 2s, although approximately 20\%\ larger.  Thus, assuming the 2p rate coefficient to be a factor of 1.2 larger, for the $n=2$ state we adopt the 2s result multiplied by a factor of $1.15$, which combines the 2s and 2p results according to their statistical weights ($2/8 + 6/8 \times 1.2$).   

\begin{figure}
\begin{center}
\resizebox{\hsize}{!}{\rotatebox{0}{\includegraphics{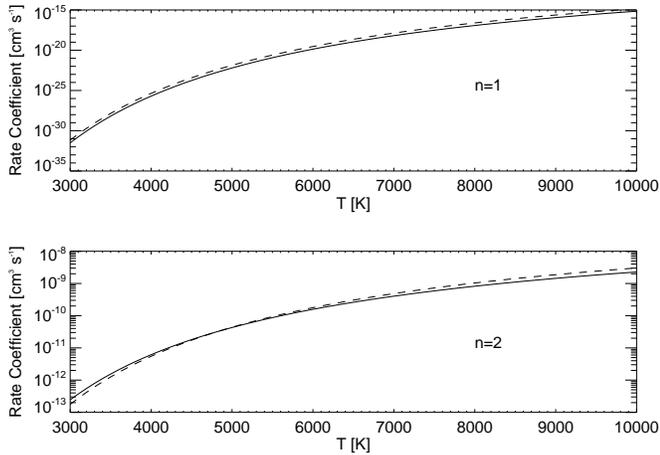}}}
\end{center}
\caption{Comparison of rate coefficients for ionisation of $n=1$ (upper panel) and $n=2$ (lower panel) levels of hydrogen.  The full lines are derived from the experimental results discussed in the text, and are used in the calculations.  The dashed lines show the result from the approximation of Vriens \&\ Smeets~(\cite{vriens80}). }
\label{fig:comp_eion}
\end{figure}

For excitation and ionisation involving higher states ($n > 7$ for excitation, $n > 2$ for ionisation), formulae from Vriens \&\ Smeets~(\cite{vriens80}) are used, as in Carlsson \& Rutten~(\cite{carlsson92}).  These analytical formulae are a semi-empirical combination of experimental and theoretical data.   Przybilla \&\ Butler~(\cite{przybilla04a}) made a comparison of available data for excitation and found the data of Percival \&\ Richards (\cite{percival78}) to be the best alternative among several well-established approximations (not including Vriens \&\ Smeets).  Noting that for the case of hydrogen the expressions of Percival \&\ Richards (\cite{percival78}) are equivalent to the expressions of Gee~et~al.~(\cite{gee76}), Vriens \&\ Smeets~(\cite{vriens80}) compare extensively with these results and find excellent agreement (see, e.g.,  Figs.~6-8 of Vriens \&\ Smeets~\cite{vriens80}), and claim a wider range of validity.  Figure~\ref{fig:comp_eion} compares the ionisation rate coefficients from Vriens \&\ Smeets~(\cite{vriens80}) with those derived from experimental data for the two lowest states discussed above.  The agreement is very good at the temperatures of interest, giving confidence that this approximate formula probably gives reasonable results for more excited states.

\paragraph{Collisional excitation and ionisation by H atoms}
\begin{eqnarray}
\mathrm{H}(n) + \mathrm{H}(\mathrm{1s}) & \rightleftharpoons & \mathrm{H}(n^\prime) + \mathrm{H}(\mathrm{1s}) \label{eqn:Hexcd} \\
                                        & \rightleftharpoons & \mathrm{H}(\mathrm{1s}) + \mathrm{H}(n^\prime) \label{eqn:Hexct} 
\end{eqnarray} 
\begin{eqnarray}
\mathrm{H}(n) + \mathrm{H}(\mathrm{1s}) & \rightleftharpoons & \mathrm{H}^+ + \mathrm{H}(\mathrm{1s}) + \mathrm{e} \label{eqn:Hiond} \\
                                        & \rightleftharpoons & \mathrm{H}(\mathrm{1s}) + \mathrm{H}^+ + \mathrm{e} \label{eqn:Hiont} 
\end{eqnarray}  
The excitation process (\mbox{$n^\prime > n$}) includes two possible channels, the direct excitation process (\ref{eqn:Hexcd}) and the excitation process involving excitation transfer (\ref{eqn:Hexct}).  Similarly, ionisation includes the direct process (\ref{eqn:Hiond}) and the channel involving an electron transfer (\ref{eqn:Hiont}).  We note also the existence of the associative ionisation process $\mathrm{H}(n) + \mathrm{H}(\mathrm{1s}) \rightleftharpoons  \mathrm{H}_2^+ + \mathrm{e}$, which is not included. 

Data for these processes are available from a number of sources.  As we will see, comparison of the data from these sources give a wide range of values for the rate coefficients, and it is not always clear which set of data are to be preferred in each instance.  We will discuss each set of data in turn, and then compare the data for some example cases.

\begin{description}

\item[\it Bates \& Lewis~(\cite{bates55}):] Data for the $n=3\rightarrow 2$ process have been calculated by Bates \& Lewis~(\cite{bates55}) using the Landau-Zener model.  These calculations consider the ionic curve crossing mechanism, which is only important for this one excitation transition in hydrogen. This process may occur only between the $n=2,3,4$ levels as only these states have avoided crossings with the ionic state potential; the $n=4$ crossing is at large internuclear distance and is thus inefficient (Bates \& Lewis~\cite{bates55}).  A fit to the rate coefficient for the reverse process is given in Table~\ref{tab:ratefits}.

\item[\it Drawin~(\cite{drawin68, drawin69}):] General analytic expressions for the cross sections and rates for both excitation and ionisation are given by Drawin (\cite{drawin68, drawin69}; see also Drawin \& Emard~\cite{drawin73}).  The expressions are derived from a semi-empirical modification of the classical Thomson formula for ionisation by electrons.  Fleischmann \& Dehmel~(\cite{fleischmann72}) have pointed out that the derivation has some apparent ``inconsistency and arbitrariness''.  For example, despite the fact that Drawin states clearly that the atomic kinetic energy $E_a$ is in the centre-of-mass frame, this is inconsistent with the stated threshold of twice the ionisation energy which would instead imply that $E_a$ is in the laboratory frame.  Furthermore, if we numerically integrate rate coefficients from Drawin's cross sections, we find that good agreement with his expression for the rate coefficient can only be obtained if we assume $E_a$ is in the laboratory frame.  Moreover, comparison with experiment for the case of ionisation (see, e.g., Fig.~3 of Kunc \& Soon~\cite{kunc91}), shows that while the formula may give reasonable results for collisions near the peak in the cross section at about $E_\mathrm{lab} \sim 10^4$~eV, the results at lower energy (100--1000~eV) are not in good agreement.  Note that for the temperatures of interest the near threshold collisions ($E_\mathrm{lab} = 2 E_\mathrm{cm} = 27$--40~eV) are of interest.   

\item[\it Soon~(\cite{soon92}):] Similarly general analytical expressions for the excitation and ionisation cross sections and rates have been derived by Soon~(\cite{soon92}), that are an extension of the work of Kunc \& Soon~(\cite{kunc91}) based on the classical impulse approximation (Gryzinski~\cite{gryzinski65}).    

Unfortunately, we find problems with the presented formulation in this case also. The variable $E$ is the impact energy, and appears to be defined as the kinetic energy in the laboratory frame, since it is stated that the relative kinetic energy in the centre-of-mass frame is given by $E_\mathrm{cm} = E/2$.  If we take the example of the ionisation case, the cross sections have a threshold in the laboratory frame of $U_n$, the ionisation potential of the target atom in the $n$th level.  This leads to a threshold energy in the centre-of-mass frame, and thus a lower limit of integration in the calculation of the rate coefficient, of $E_\mathrm{min}=U_n/2$.  The threshold in the centre-of-mass frame should be the ionisation energy $U_n$, and in the laboratory frame should be $2 U_n$.  The laboratory frame threshold is larger since half of the kinetic energy is, due to conservation of momentum, unavailable, since it is locked in the centre-of-mass motion. 

Furthermore, we experienced difficulties in reproducing the results for the rate coefficients.  While we were able to perfectly reproduce the cross sections of Soon from the expressions, when we calculated rate coefficients by direct numerical integration over the Maxwellian distribution of velocities, we encountered problems.  Following exactly the formulae presented in Soon's paper, for the cross sections and the rate coefficient integrations, (i.e. making no attempt to correct the apparent problem with the thresholds, thus adopting $E=E_\mathrm{lab}$ and $E_\mathrm{min}=U_n/2$) we found significant differences between our results and the directly integrated results plotted in Soon for the rate coefficients at low temperature ($T\la 10^5$~K) and low principal quantum number ($n\la 10$).  However, the results are in good agreement for high $T$ and high $n$.  Similar discrepancies were seen with the results from Soon's analytical expressions for the rate coefficients.  To investigate further, we analytically integrated the low energy cross section expression for ionisation (Soon's eqn.~10), this regime dominating at the temperatures of interest, with the aid of the symbolic computing package \texttt{Mathematica}.  The results were in perfect agreement with our numerical results.  At the temperatures of interest in this work ($T\sim 5000$~K), the differences between the results are significant, with our rate coefficients often 1--2 orders of magnitude smaller than those of Soon in this temperature regime.  We note that this problem cannot be explained by instead adopting $E$ as the kinetic energy in the centre-of-mass frame $E_\mathrm{cm}$, or a simple correction of the threshold in the integrations, as both these changes further reduce the rate coefficients at the temperatures of interest, since the threshold is shifted to higher energies.

\item[\it Mihajlov et~al.(\cite{mihajlov96,mihajlov04,mihajlov05}):] For the excitation from levels \mbox{$4 \le n \le 10$} where \mbox{$n^\prime -n \le 5$}, rate coefficients from Mihajlov et~al.~(\cite{mihajlov04,mihajlov05}) are available.   For the ionisation of levels \mbox{$4 \le n \le 10$}, rate coefficents are available from Mihajlov~et~al.~(\cite{mihajlov96}).  These semi-classical calculations are based on the quasi-resonant energy transfer mechanism (e.g. Janev \& Mihajlov~\cite{janev79}) and account for processes (\ref{eqn:Hexcd}--\ref{eqn:Hiont}), including those involving electron transfer.  

\end{description}

\begin{figure}
\begin{center}
\resizebox{\hsize}{!}{\rotatebox{0}{\includegraphics{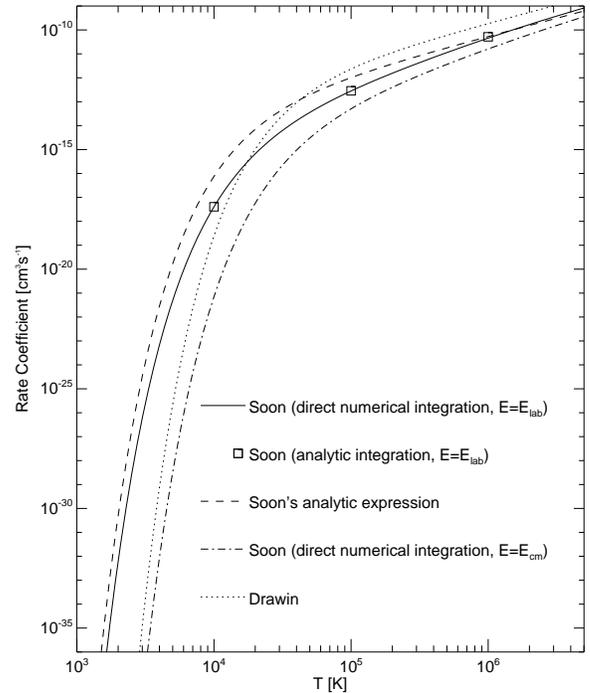}}}
\end{center}
\caption{Comparison of rate coefficients for collisional ionisation of H from the ground level due to ground state H atoms, $\mathrm{H}(1s)+\mathrm{H}(1s) \rightarrow \mathrm{H}(1s)+\mathrm{H}^++e$.  The full line is our result from direct numerical integration of the cross sections of Soon~(\cite{soon92}), where $E=E_\mathrm{lab}$.  The dot-dashed line shows the result if we instead assume $E$ is the kinetic energy in the centre-of-mass frame $E=E_\mathrm{cm}$, rather than in the laboratory frame.  The squares show results from analytic integration of Soon's eqn.~10, the low energy cross section expression.  The dashed line shows the result from Soon's analytic expression for the ionisation rate coefficient. 
The dotted line shows the result from the analytic expression of Drawin~(\cite{drawin68, drawin69}).}
\label{fig:comp_Hion}
\end{figure}

Comparison of the various data sets, where they overlap, is instructive.  We begin with the case of ionisation from the ground level, as this also illustrates the problems with the rate coefficients from Soon.  A comparison of the rate coefficients from different sources is given in Fig.~\ref{fig:comp_Hion}.  First, we note the apparent problems with the Soon cross sections and rate coefficients.  At $T \la 10^5$~K, our numerically integrated rate coefficients are typically about an order of magnitude smaller than those given by Soon's analytic approximation, and those numerically integrated by Soon (not shown, see Fig.~11 of Soon's paper).  It is also demonstrated that adopting $E=E_\mathrm{cm}$ worsens the disagreement.  We note also that at the temperatures of interest for the photosphere ($10^3 < T < 10^4$~K), the Drawin formula gives much lower rate coefficients than Soon.  

Comparison of the cross sections with experimental results available in the energy range $E\approx 10^2$--$10^5$~eV (see Fig.~1 of Soon~\cite{soon92}, and Fig.~3 of Kunc \& Soon~\cite{kunc91}), shows the Soon cross sections are in much better agreement with experiments than those of Drawin.  This conclusion would not be greatly affected by assuming $E=E_\mathrm{cm}$, even though the comparison with experiment would be worsened for the Soon cross sections.  However, the rate coefficients at the temperatures of interest are severely affected by such a change, since the cross sections near the threshold ($E_\mathrm{cm} = E_\mathrm{lab}/2 \ga 13.6$~eV) dominate.  In fact, the Soon rate coefficients ($E=E_\mathrm{lab}$) are only larger than those of Drawin due to the incorrect placement of the threshold.  If we correct the threshold of the Soon cross sections by assuming $E=E_\mathrm{cm}$, then the rate coefficients become even smaller than those of Drawin (see Fig.~\ref{fig:comp_Hion}).  We note, it is dangerous to infer too much about the accuracy of the cross sections near the threshold from the comparison with experiments at much higher energies.  

\begin{figure}
\begin{center}
\resizebox{\hsize}{!}{\rotatebox{0}{\includegraphics{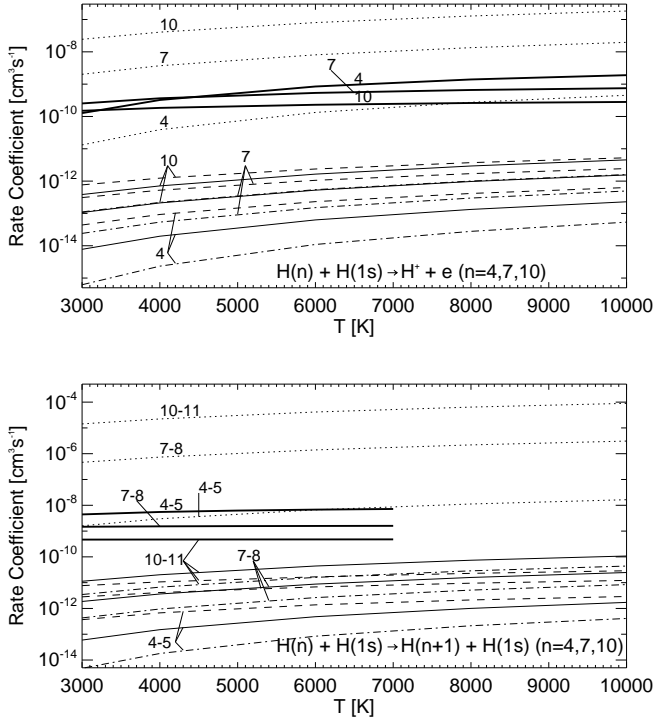}}}
\end{center}
\caption{Comparison of collisional ionisation (upper panel) and excitation (lower panel) rate coefficients due to ground state H atoms for selected excited states. The data from Mihajlov~et~al.~(\cite{mihajlov96}, \cite{mihajlov04}) are plotted as thick full lines.  The remaining lines follow the scheme used in Fig.~\ref{fig:comp_Hion}, i.e., the numerically integrated data from Soon~(\cite{soon92}, $E=E_\mathrm{lab}$) are plotted as normal full lines.  The dot-dashed line shows the result if we instead assume $E=E_\mathrm{cm}$.  The data from Soon's analytic expressions are plotted as dashed lines, and the data from Drawin~(\cite{drawin68, drawin69}) are plotted as dotted lines.  The dot-dashed line for ionisation from $n=7$ is under the full line for $n=10$. }
\label{fig:comp_Hcoll}
\end{figure}

Comparisons for ionisation from some selected excited states are plotted in the upper panel of Fig.~\ref{fig:comp_Hcoll}.  This has the advantage of enabling comparison with the results of Mihajlov~et~al.~(\cite{mihajlov96}), which we judge to be much more reliable than the classical theories.  An immediately noticeable result is that for the temperatures of interest, the various sources do not follow the same relative behaviour as for ionisation from the ground state.  For excited states, it is clear that the Drawin recipe is typically several orders of magnitude larger than the Soon results, irrespective of chosen integration recipe.  In general (except for the low-lying states), the Mihajlov~et~al.\ results lie somewhere in between, smaller than Drawin, yet larger than Soon.  

The results for excitation, plotted in the lower panel of Fig.~\ref{fig:comp_Hcoll}, are similar. One may also compare the rate coefficients for $n=2\rightarrow 3$ computed from the Bates \& Lewis~(\cite{bates55}) data via the detailed balance relation, with those from the general classical recipes.  However, one must realise that the comparison is of limited value since the curve crossing mechanism at work here is not reflected in the classical models.  Any agreement would be purely coincidental.  If we compare results at 8000~K, the Drawin formula gives a rate coefficent over an order of magnitude larger than the Bates \& Lewis value.  The analytical formula of Soon gives a value a factor of two smaller, while numerical integrations with $E=E_\mathrm{lab}$ and $E=E_\mathrm{cm}$ are roughly one and three orders of magnitude smaller, respectively.

These comparisons clearly demonstrate the present uncertainties on the rate coefficients due to hydrogen collisions in many cases.  It is our judgement that the rate coefficients of Mihajlov~et~al.~(\cite{mihajlov96,mihajlov04,mihajlov05}), and Bates \& Lewis~(\cite{bates55}), are to be preferred over the two classical recipes.  Thus, these data are adopted wherever possible, namely for the ionisation from \mbox{$4 \le n \le 10$} and excitation for the same levels where \mbox{$n^\prime -n \le 5$}, and for the case of the $n=2\rightarrow 3$ transition.  Unfortunately, given the discussed problems with the two classical recipes, we have no real reason to prefer one over the others.  For this reason we have chosen to perform calculations for four models.  We define the following four recipes for calculating the remainder of the required rate coefficients:
\begin{description}
\item[{\it HD}] The Drawin rate coefficients are used.
\item[{\it HSa}] Soon's analytical expressions (his equations 15 and 16) are used.
\item[{\it HSn}] Rate coefficients obtained from numerical integration of Soon's cross section expressions, assuming $E=E_\mathrm{lab}$, are used.
\item[{\it HSnc}] Rate coefficients obtained from numerical integration of Soon's cross section expressions, assuming $E=E_\mathrm{cm}$, are used.
\end{description}
We will refer to these recipes in what follows.  Our main aim will be to determine how the uncertainties in the atomic data translate to uncertainties in the non-LTE modelling.  While we hope this may give us some guidance, we should point out that there is no reason to believe that these recipes bracket the true values.

\paragraph{Mutual neutralisation}
\begin{equation}
\mathrm{H}^+ + \mathrm{H}^- \rightleftharpoons  \mathrm{H}(n=2,3) + \mathrm{H}(\mathrm{1s}) 
\label{eqn:chex}
\end{equation}
This charge exchange process was included employing the cross sections of Fussen \& Kubach (\cite{fussen86}), which have been computed using a quantum close coupling treatment.  This process may occur only into the $n=2,3,4$ levels as only these states have avoided crossings with the ionic state potential.  As discussed above, the $n=4$ crossing is at large internuclear distance, and is thus inefficient.  The $n=3$ level, in particular, has a significant charge exchange cross section.  The rate coefficients for the mutual neutralisation were determined from these cross sections, and fits to these results are given in Table~\ref{tab:ratefits}.  Note, as H$^-$ is not included in our model atom, in calculating the collision rates for this process we have assumed LTE populations for H$^-$.  As this process could only conceivably be important in the deep photosphere where H$^-$ is in LTE (see discussion in \S\ref{sect:intro}) this approximation should be adequate.

\paragraph{Penning ionisation involving H(n=2)}
\begin{equation}
\mathrm{H}(n=2) + \mathrm{H}(n=2) \rightleftharpoons \mathrm{H}(\mathrm{1s}) + \mathrm{H}^+ + \mathrm{e}.
\label{eqn:penning}
\end{equation}        
The quantal calculations of Bates et~al.~(\cite{bates67}) were employed, specifically $\langle \sigma \mathrm{v} \rangle = 2.1\times 10^{-8} T^{-1/6}$~cm$^3$~s$^{-1}$ for the forward reaction.  Such collisions are expected to be rare, but we have modelled this process anyway as it is easily included. 

\begin{table}
\caption{Polynomial fits to the rate coefficient for collision processes.  The rate coefficients $\langle \sigma \mathrm{v} \rangle$ are given by equation~\ref{eqn:ratefit} in units of cm$^3$~s$^{-1}$ for the listed process. 
 The fits were made to results over the range $1000 < T < 20000$~K.}
\label{tab:ratefits}
\begin{center}
\begin{tabular}{ccccc}
\hline
\hline
$n$ or $nl$ & $a_1$ & $a_2$ & $a_3$ & $a_4$  \\
\hline
 & & & & \\
\multicolumn{5}{c}{\underline{$\mathrm{H}(nl) + \mathrm{e} \rightarrow \mathrm{H}^+ + 2\mathrm{e} $}} \\
1s  & $-2324.8$ & $+1642.6$ & $-391.96$ & $+31.41$ \\  
2s  &  $-582.7$ &  $+409.9$ &  $-98.28$ &  $+7.92$ \\   
 & & & & \\
\multicolumn{5}{c}{\underline{$\mathrm{H}(n=2) + \mathrm{H}(\mathrm{1s}) \rightarrow \mathrm{H}(n=3) + \mathrm{H}(\mathrm{1s})$}} \\
2   & $-341.5$ & $+233.9$ & $-55.46$ & $+4.388$ \\  
 & & & & \\
\multicolumn{5}{c}{\underline{$\mathrm{H}^+ + \mathrm{H}^- \rightarrow \mathrm{H}(n) + \mathrm{H}(\mathrm{1s})$}} \\
2   & $-19.08$ & $+8.450$ & $-2.500$ & $+0.249$ \\  
3   & $-6.074$ & $+0.048$ & $-0.222$ & $+0.031$ \\  
\hline
\end{tabular}
\end{center}
\end{table}

\subsubsection{LTE model atom}

It will be of interest to compare the results of our non-LTE calculations with the LTE results.  To consistently simulate LTE conditions, LTE results are calculated using the same codes using a model atom with extremely large electron collision rates such that all departure coefficients are unity. 

\subsection{Model stellar atmospheres}

Statistical equilibrium calculations as described above were carried out for a semi-empirical solar atmosphere model including a chromosphere, and a number of theoretical stellar atmosphere models.  Calculations were performed for the semi-empirical model of the quiet Sun of Maltby et~al.~(\cite{maltby86}), hereafter referred to as the MACKKL model.  The theoretical models are computed under the assumptions of 1D plane-parallel geometry and LTE using the \texttt{MARCS} code (Gustafsson et~al.~\cite{gustafsson75}, Asplund et~al.~\cite{asplund97}).  Scaled solar abundances are used based on solar values from Grevesse \&\ Sauval~(\cite{grevesse98}), except for C which is taken from Allende Prieto~et~al.~(\cite{allende02}).  Convection is treated under the mixing-length theory with the parameters $\alpha=1.5$ and $y=3/4\pi^2$, which are the default values for \texttt{MARCS}.  See, for example, Henyey~et~al.~(\cite{henyey65}) for details.  As mentioned earlier, various authors have suggested revisions of these parameters, particularly $\alpha$, to better reproduce Balmer line profiles in solar-type stars.  However, this investigation is a study of line formation processes and this is not important here.   
                                                            
\section{Results}
\label{sect:nonlte_results}

We begin our study of Balmer line formation with the case of the Sun using the MACKKL semi-empirical solar model with chromosphere.  This permits us to investigate the effect of various collision processes in both the photosphere and the chromosphere, where the wings and line cores, respectively, predominantly form.   It is worth emphasising that we study line formation within a given model atmosphere, and any feedback of our different model atom on the semi-empirical temperature structure is neglected (see, e.g., Fontenla~et~al.~\cite{fontenla06}).  Finally, we investigate how the results for the line wings translate to other late-type stars using theoretical model atmospheres with no chromospheres.

In order to gauge the effect and importance of the various collision processes, we performed calculations separating the various collision processes.  As collisions with electrons are well understood to be important, we performed a calculation including only electron collisions.  Subsequent calculations were then performed including additional collision processes.  First, mutual neutralisation and Penning ionisation processes were added.  As we will see, these processes have negligible effects on the results.  We then added collisions with hydrogen atoms, following one of the four recipes discussed above, labelled HD, HSa, HSn, and HSnc.  For brevity, in the following we will refer to these models by these labels, noting that it is always implied that electron collisions, mutual neutralisation, and Penning ionisation processes are included.

\subsection{The solar atmosphere}

\begin{figure}
\begin{center}
\resizebox{\hsize}{!}{\rotatebox{0}{\includegraphics{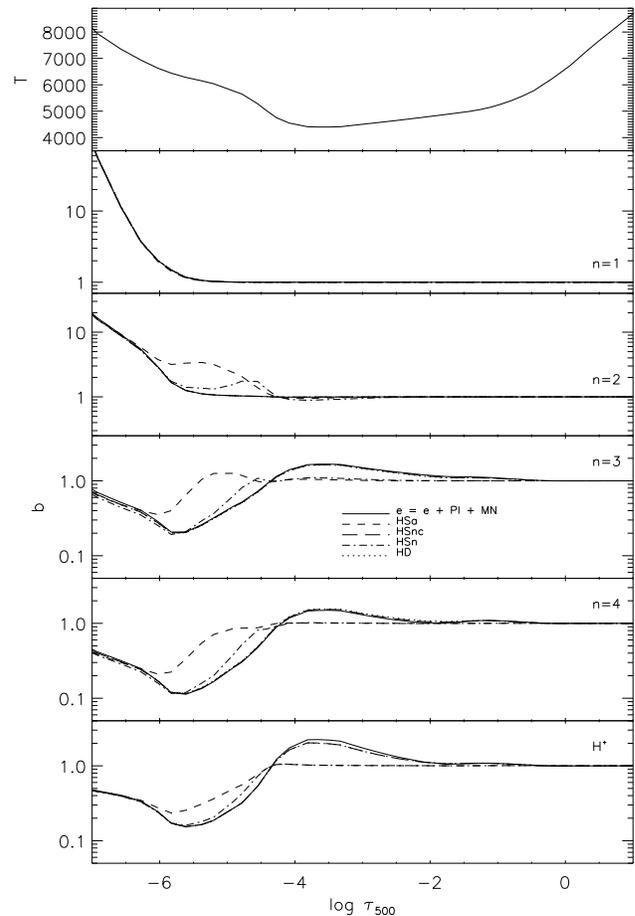}}}
\end{center}
\caption{Results for the departure coefficients $b$ of the four lowest levels of H and for H$^+$ with the logarithm of continuum optical depth at 500 nm, $\log \tau_{500}$, from calculations including different collision processes for the MACKKL semi-empirical solar model. In the upper panel the temperature stratification is shown for reference. The full line shows the result where only electron collisions are included.  The difference if Penning ionisation (PI) and mutual neutralisation (MN) processes are included, in addition to the electron collisions, would not be noticeable on this plot.  The additional lines show the results if all collision processes are included, where the various different recipes for the hydrogen collisions are used. }
\label{fig:mackkl_ncoll}
\end{figure}

\begin{figure}
\begin{center}
\resizebox{\hsize}{!}{\rotatebox{0}{\includegraphics{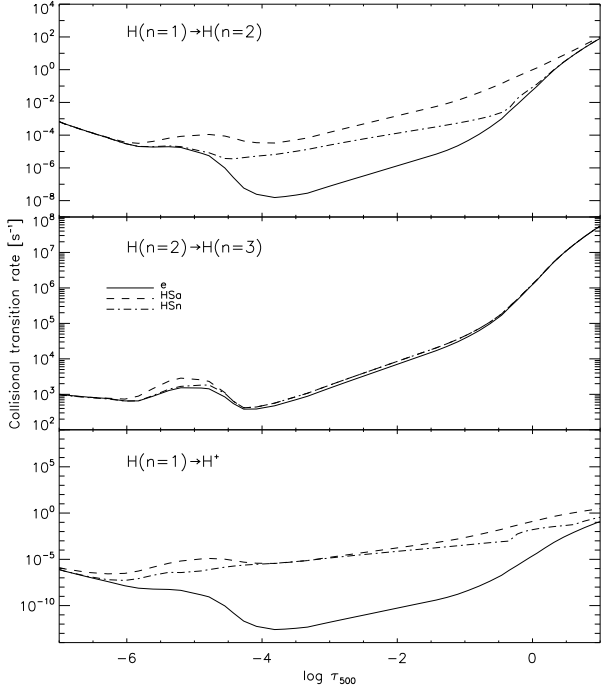}}}
\end{center}
\caption{Results for the total collision rates between selected levels with the logarithm of continuum optical depth at 500 nm, $\log \tau_{500}$, from calculations including different collision processes for the MACKKL semi-empirical solar model.  The full line shows the result where only electrons are included.  The dashed and dash-dot lines show the results from the HSa and HSn models, respectively.  The HSnc and HD models are not shown as they show only minor differences from the model including only electron collisions.  Penning ionisation and mutual neutralisation processes do not affect these transitions.  Note, the $n=2\rightarrow 3$ transition is in fact calculated using the Bates \& Lewis data in all cases.  The small differences arise due the differences in populations.}
\label{fig:mackkl_collrates}
\end{figure}

Figure~\ref{fig:mackkl_ncoll} presents the departure coefficient results from these calculations for the MACKKL model.  To aid in interpretation of these results, Fig.~\ref{fig:mackkl_collrates} plots the total collision rates for selected important transitions.  First, we point out the result mentioned above, the models including electrons only, and electrons plus mutual neutralisation and Penning ionisation processes, are essentially identical.   Penning ionisation involving H$(n=2)$, process~(\ref{eqn:penning}), was found to have practically no effect on the statistical equilibrium, as would be expected due to the rarity of such collisions.  The mutual neutralisation charge exchange process~(\ref{eqn:chex}), couples the $n=2$ and $n=3$ states slightly more strongly to the ionised state, but has only barely noticeable effects.  Secondly, we notice that the results using the HD and HSnc models are not significantly different from those with electrons only.  Finally, we see that the results from the HSa and HSn models do show significant differences from those where only electrons are included.  As such, in the following discussions we focus predominant on the HSa and HSn models since they show the most significant changes; however, this should not be interpreted as favouring these models.

\begin{figure}
\begin{center}
\resizebox{\hsize}{!}{\rotatebox{0}{\includegraphics{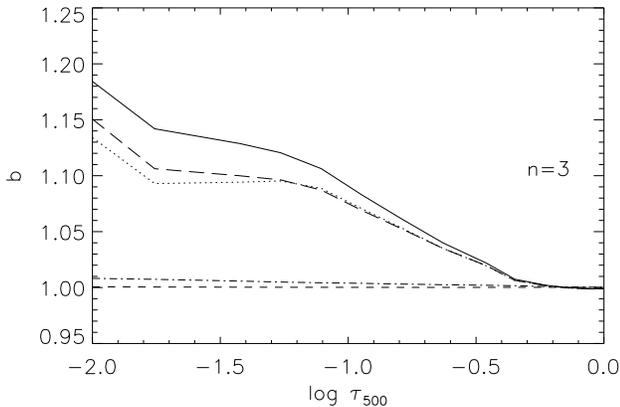}}}
\end{center}
\caption{Closer view of the departure coefficients $b$ of the $n=3$ state with the logarithm of continuum optical depth at 500 nm, $\log \tau_{500}$, from calculations including different collision processes for the MACKKL semi-empirical solar model, in the photosphere.  Lines have the same meanings as in Fig.~\ref{fig:mackkl_ncoll}.}
\label{fig:mackkl_ncoll2}
\end{figure}

In the photosphere, below the temperature minimum ($\log \tau \ga -3$), the $n=1,2$ states attain LTE populations even with only electron collisions included, but the higher states and the ionised state show overpopulation in the upper photosphere.  To show the situation more clearly, Fig.~\ref{fig:mackkl_ncoll2} shows the departure coefficients for the photosphere for the $n=3$ state, which is representative of the effect in the higher states.  The departures practically disappear when excitation and ionisation by collisions with ground state hydrogen atoms, processes~(\ref{eqn:Hexcd}--\ref{eqn:Hiont}), are included following recipes HSn or HSa.  In fact, test calculations showed that the overpopulations of the H$^+$ and $n\ge 3$ states in this region of the atmosphere are removed by inclusion of only the ionisation by H atom collision of the $n=1$ state.  As can be seen in Fig.~\ref{fig:mackkl_collrates}, the collision rate for this process increases by several orders of magnitude, and thus the $n=1$ and H$^+$ states are brought close to detailed balance bringing H$^+$ to LTE, with the effect trickling down to the upper states via their already strong coupling to the continuum.

\begin{figure}
\begin{center}
\resizebox{\hsize}{!}{\rotatebox{0}{\includegraphics{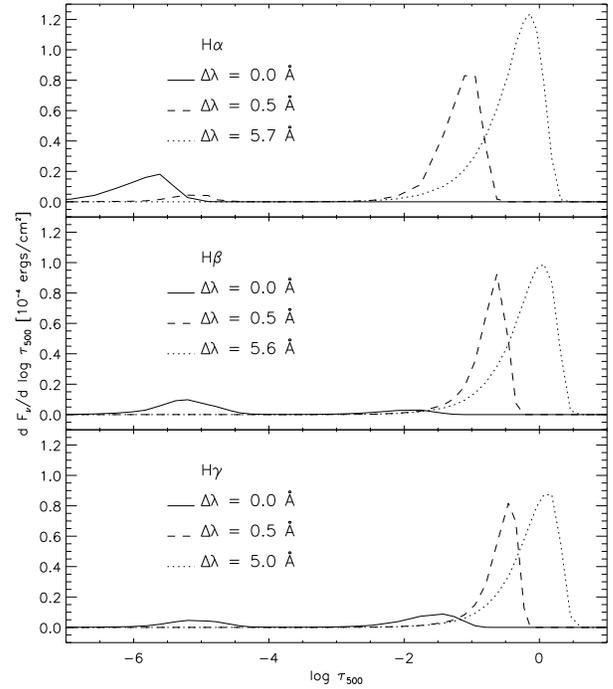}}}
\end{center}
\caption{Contribution functions to the flux with the logarithm of continuum optical depth at 500 nm, $\log \tau_{500}$, for H$\alpha$, H$\beta$, and H$\gamma$, at three detunings $\Delta \lambda \sim 0, 0.5, 5$~\AA\ chosen to show the behaviours in the core, core-wing transition region, and wing respectively.  These were calculated using the HD model, but do not significantly depend on this choice.}
\label{fig:mackkl_cntrb}
\end{figure}

We now turn our attention to the effect on the Balmer line profiles. It is useful for our discussion to have a quantitative idea of where the lines form, and contribution functions for the core, core-wing transition and wing for the lower three Balmer lines have been plotted in Fig.~\ref{fig:mackkl_cntrb} for reference.

\begin{figure}
\begin{center}
\resizebox{\hsize}{!}{\rotatebox{0}{\includegraphics{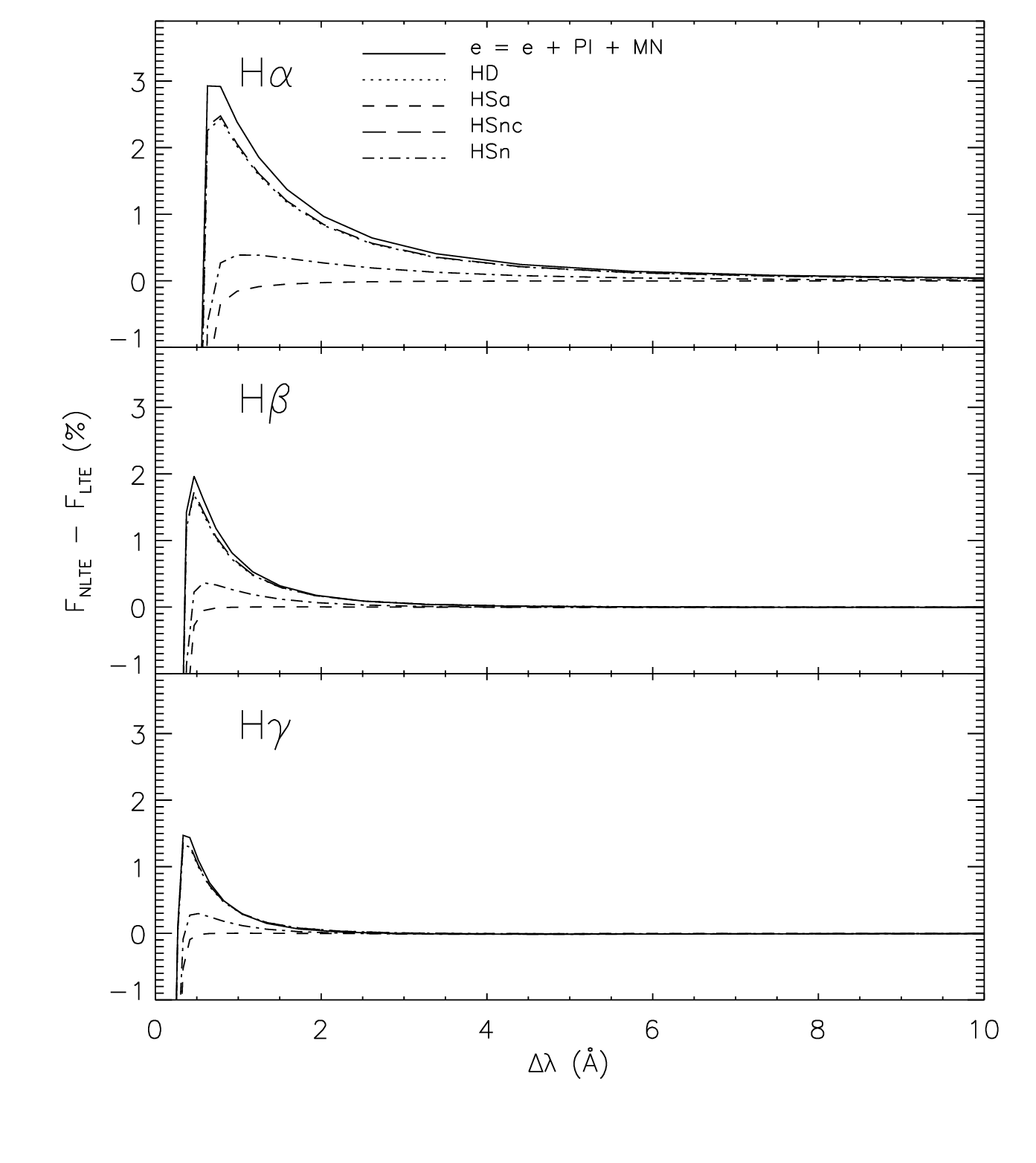}}}
\end{center}
\caption{The effect of different collision processes on the wings of the lower Balmer lines in the semi-empirical MACKKL model for the solar atmosphere.  The difference between the non-LTE line profile $F_\mathrm{NLTE}$ and the LTE profile $F_\mathrm{LTE}$ is plotted in units of per cent of continuum flux.   The full lines show the results when only electron collisions are included.  The remaining lines show the results when all collisional processes are included, where various different recipes for the hydrogen collisions are used.}
\label{fig:mackkl_coll}
\end{figure}

First, we focus our attention on the line wings which form deep in the photosphere at $\log \tau_{500} > -2$ (see Fig.~\ref{fig:mackkl_cntrb}). In Fig.~\ref{fig:mackkl_coll} the differences between the LTE profiles and non-LTE profiles in the line wings for H$\alpha$, H$\beta$, and H$\gamma$ are shown.  First, we have the notable result, that the lines are not well approximated by LTE if only electron collisions are included.  Secondly, we see that even with hydrogen collisions added, the result depends on which recipe for the rate coefficients is used.  The results for the HD and HSnc models are very similar, and show only a small difference from the case with only electrons.  For the HSn model, the line profiles are close to the LTE profiles in the wings, but still some small differences exist.  For the HSa model, the line wings are in good agreement with those predicted in LTE, and are indeed formed in LTE as the departure coefficients and $S_\nu/B_\nu$ are unity in the region where the line wings form.  Higher Balmer lines show similar results.

\begin{figure}
\begin{center}
\resizebox{\hsize}{!}{\rotatebox{0}{\includegraphics{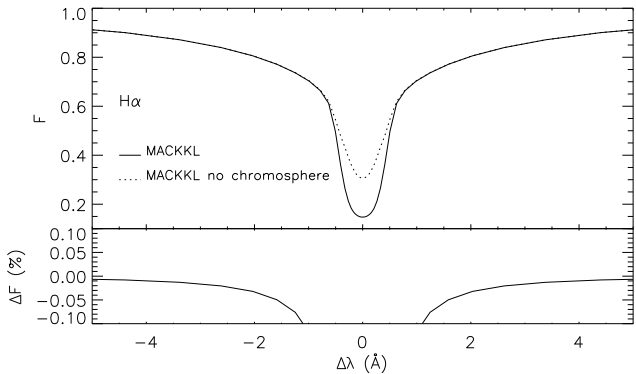}}}
\end{center}
\caption{Comparison of H$\alpha$ calculated for the MACKKL model both with and without chromosphere.  The calculations were done for the HSa model.}
\label{fig:mackkl_nochromo}
\end{figure}

In Fig.~\ref{fig:mackkl_coll}, it is notable that even in the case of the HSa model, which attains LTE to high accuracy in the line wings, in the inner wing region of H$\alpha$ around $1 \la \Delta \lambda \la 2$~\AA\ there are very small ($\la 0.1$~\%) departures from the LTE profile.  This is due to small contributions from the chromosphere to this part of the line, as demonstrated in Fig.~\ref{fig:mackkl_cntrb} showing the small contributions from $\Delta \lambda = 0.5$.  This is shown more clearly in Fig.~\ref{fig:mackkl_nochromo}, which compares the calculation for H$\alpha$ using the full MACKKL model with chromosphere, with a calculation employing the MACKKL model where layers above the temperature minimum ($\log \tau_{500} < -3.6$) are removed.  The far wings show excellent agreement while small differences are see in the inner wings.  The question if there is an effect of the chromosphere on the formation of the line wings is of importance.  The use of the Balmer line wings as effective temperature indicators depends on the fact that they are formed in the photosphere, which can be modelled theoretically from the basic stellar parameters (usually $T_\mathrm{eff}$, $\log g$ and chemical composition).  While the effect seen here in the solar case is extremely small, it does perhaps warn of possible effects on the inner wing region in more active stars.  As an aside, we note that Przybilla \& Butler~(\cite{przybilla04b}) found that the H$\alpha$ line profiles, including the cores, were almost identical when computed in this manner (MACKKL with and without chromosphere), though with quite different line formation physics in each case.  We reproduced this result if only electron collisions are included (or models HD or HSnc); however, as seen in Fig.~\ref{fig:mackkl_nochromo}, this result does not occur if the HSa model is used (or HSn). 

\begin{figure}
\begin{center}
\resizebox{\hsize}{!}{\rotatebox{0}{\includegraphics{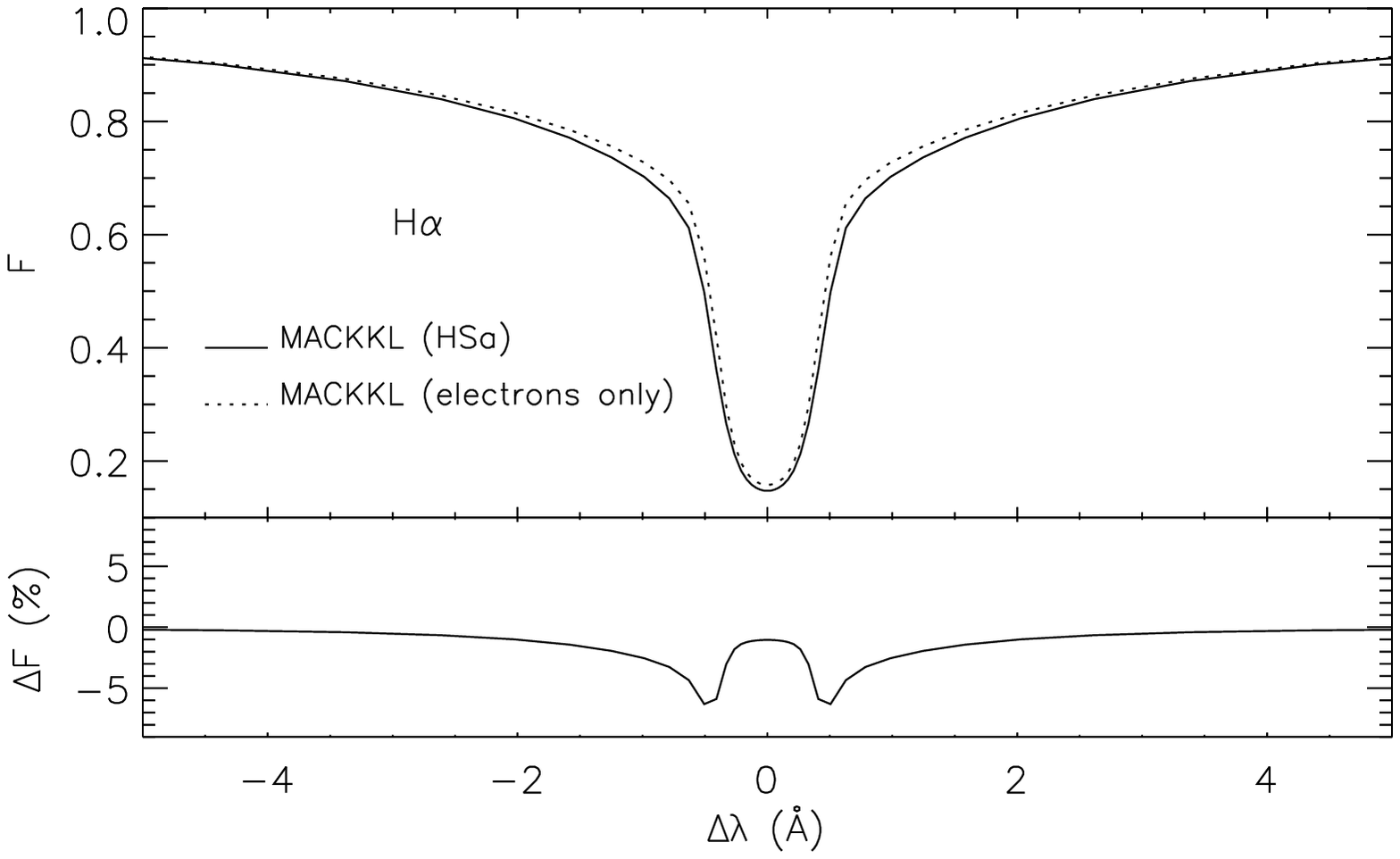}}}
\resizebox{\hsize}{!}{\rotatebox{0}{\includegraphics{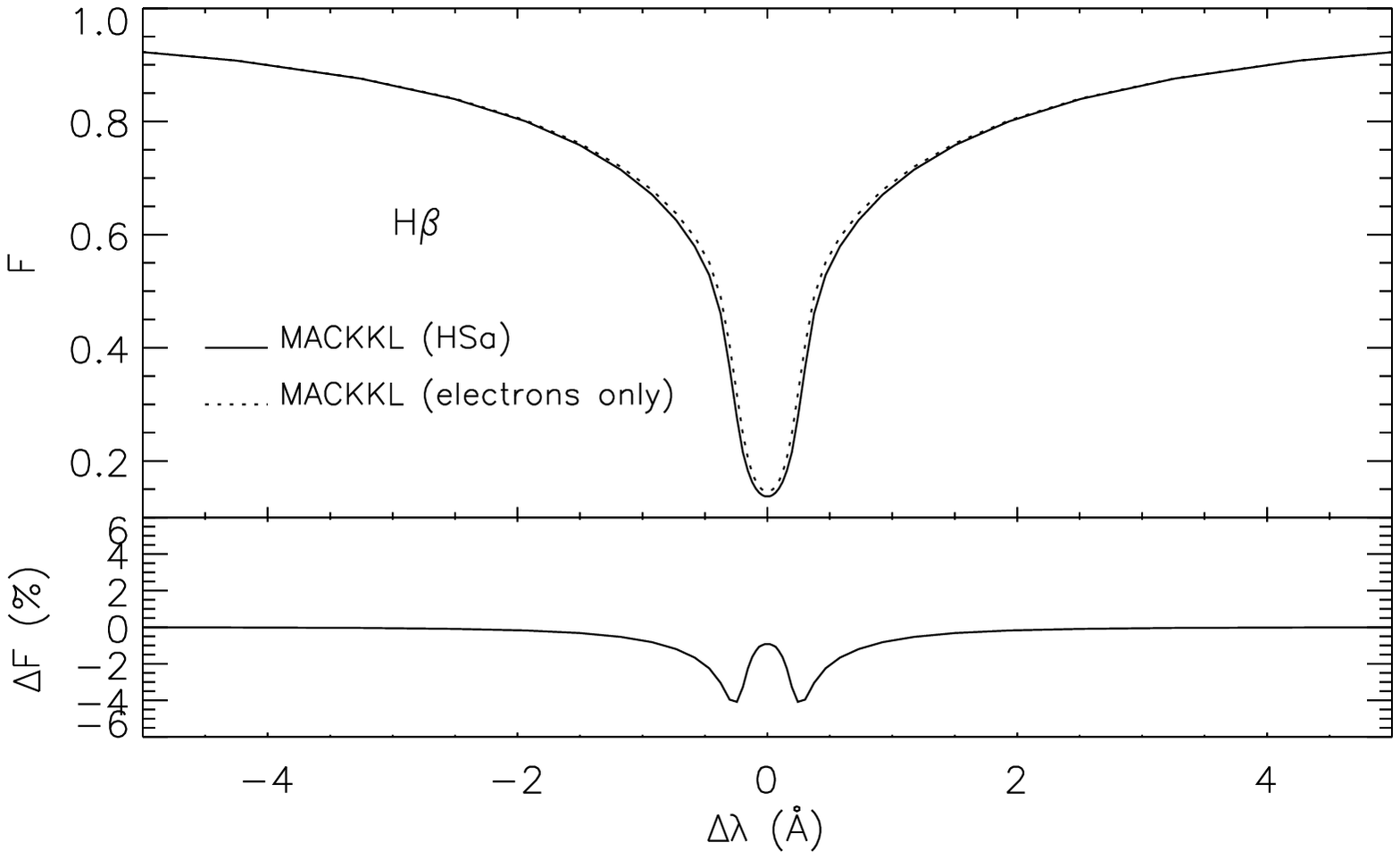}}}
\end{center}
\caption{Results for the H$\alpha$ and H$\beta$ line cores in the MACKKL model, with electrons only or the HSa recipe.}
\label{fig:mackkl_coll2}
\end{figure}

We now turn our attention to the line core.   As shown in Fig.~\ref{fig:mackkl_cntrb}, the lower Balmer line cores form predominantly in the chromosphere, where deviations from LTE are expected as the density is low and collisions are inefficient in thermalising the gas.  We note that the assumption of a static model and statistical equilibrium are known to be incorrect in the chromosphere (e.g.\ Carlsson \& Stein~\cite{carlsson02}; Leenaarts \& Wedermeyer~\cite{leenaarts06}, and references therein).  However, such a calculation may indicate if hydrogen collision processes could conceivably be important in the chromosphere, and in the formation of the line core.  Compared with when electrons only are considered, the HD and HSnc models have basically no effect on the line core.  When the HSa or HSn models are used the effect on the predicted line cores is small, of the order of a few per cent for H$\alpha$ and H$\beta$.  Figure~\ref{fig:mackkl_coll2} shows that the effect on the predicted line cores for H$\alpha$ and H$\beta$ for the case of the HSa model; the HSn results are similar.

\subsection{Late-type stellar atmospheres}
\label{sect:late}

\begin{figure}
\begin{center}
\resizebox{\hsize}{!}{\rotatebox{0}{\includegraphics{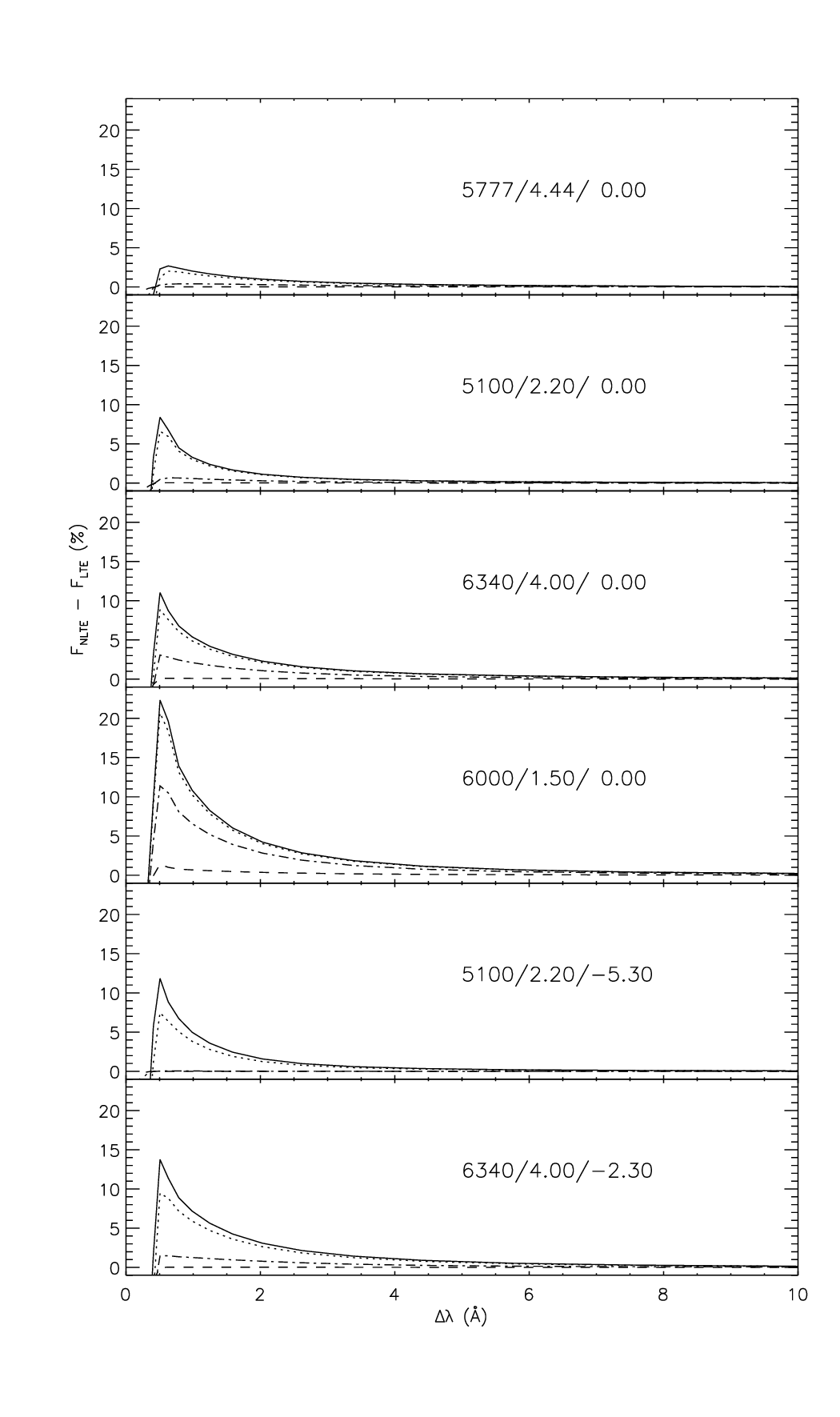}}}
\end{center}
\caption{The effect of different collision models on the wings of H$\alpha$ for various \texttt{MARCS} models.  The difference between the non-LTE line profile $F_\mathrm{NLTE}$ and the LTE profile $F_\mathrm{LTE}$ is plotted in units of per cent of continuum flux.  The lines have the same meanings as in Fig.~\ref{fig:mackkl_coll}, but the long-dashed line (HSNc model) is not plotted, as it is practically coincident with the dotted line (HD model) in all cases. Each plot is labelled by the model parameters $T_\mathrm{eff}/\log g/\mathrm{[Fe/H]}$. }
\label{fig:other}
\end{figure}

We now investigate how these results for the solar atmosphere, particularly for the Balmer line wings, translate to other late-type stars, especially the cases of metal-poor and giant stars where collisions may be less efficient.    Calculations were performed for six theoretical models covering a range of different late-type stellar atmospheres, which are summarised in Table~\ref{tab:models}.  The results for the H$\alpha$ profile are given in Fig.~\ref{fig:other}.   For reference, we included a MARCS solar model, and the results for the line wings are seen to be similar to those for the MACKKL model.  We also calculated for solar metallicity giant, turnoff star, and supergiant models.  The results are of a similar form to the solar case, although the magnitude of the differences from the LTE case for the electrons only, HD, HSn, and HSnc models increases.  The HSa model attains LTE in the line wings in all three cases.   Calculations were also performed for two metal-poor models.  The first is an extremely metal-poor giant star corresponding to the case of HE~0107$-$5240, and the second a metal-poor turnoff star corresponding to the typical example HD~84937.  The results are again similar, with the magnitude of the differences from the LTE case for the models, where they exist, being different to the corresponding solar metallicity case.  The electrons only and HD model cases show increased differences from LTE, while the HSn model results show decreased differences.  Once again, the HSa models attain LTE in the line wings.

\begin{table*}
\tabcolsep 1.5mm
\begin{center}
\caption{Details for theoretical models used.  Unless otherwise stated, models used scaled solar abundances as described in the text.  }
\label{tab:models}
\begin{tabular}{lrrrl}
\hline
\hline
Model                          & $T_\mathrm{eff}$ & $\log g$ & [Fe/H] & notes \\
                               & [K]              &[dex]     & [dex]  &       \\
\hline
Solar model                    & 5777             & 4.44     & 0.0    &       \\
Solar metallicity giant        & 5100             & 2.20     & 0.0    &       \\
Solar metallicity turnoff star & 6340             & 4.00     & 0.0    &       \\
Solar metallicity F supergiant & 6000             & 1.50     & 0.0    &       \\
Very metal-poor giant          & 5100             & 2.20     & $-5.3$ & $\alpha$-enhanced (+0.4~dex), $\mathrm{[C/Fe]} = 4.0$, $\mathrm{[N/Fe]} = \mathrm{[O/Fe]} = 2.3$       \\ 
Metal-poor turnoff star        & 6340             & 4.00     & $-2.3$ & $\alpha$-enhanced (+0.4~dex)       \\       
\hline
\end{tabular}
\end{center}
\end{table*}

\section{Discussion}
\label{sect:discussion}

As mentioned in Sect.~\ref{sect:intro}, in most past work on the statistical equilibrium of hydrogen in stellar atmospheres, only collisional processes involving electrons have been included.  This is based on the fact that at the thermal velocities of interest electron collisions are expected to be non-adiabatic and have substantial cross sections, while collisions with heavier particles are expected to be nearly adiabatic and thus have small cross sections.  In the photosphere of solar-type stars, neutral hydrogen atoms typically outnumber electrons by a factor of $10^4$, and can be orders of magnitude greater still in metal-poor stars.  For example, for the HSa and HSn models, the rate coefficients at 5000~K for the $n=1\rightarrow 2$ transition due to electrons are only of the order 15 and 2500 times greater than those due to hydrogen collisions, respectively, and thus hydrogen collisions may even be dominant in certain regions as seen in Fig.~\ref{fig:mackkl_collrates}.  For the rate coefficient for the $n=2\rightarrow 3$ transition the difference is of the order $10^6$, and thus hydrogen collisions start to become important for this transition in metal-poor stars.  Thus, sheer weight of numbers could lead to the importance of hydrogen atoms collisions, which has been seen in the presented calculations.  

It is appropriate to note here, that the above arguments speak against the possibility of inelastic collisions with protons being of any significance.  As collisions with protons are expected to be adiabatic and the proton abundance will be less than the electron abundance, such processes should always be insignificant with respect to electron collisions.  However, one must also note the limited validity of the Massey criterion.  The adiabatic approximation breaks down in the case of small energy separations, such as occurs at avoided crossings of potential energy curves (e.g.~Bohm~\cite{bohm}, page 505).

We have shown that mutual neutralisation processes and Penning ionisation are unimportant in the statistical equilibrium of hydrogen in late-type atmospheres.
Unfortunately, from the results presented here the role of collisional excitation and ionisation by hydrogen is unclear.  Two models, HD and HSnc, predict that hydrogen collisions are relatively unimportant.  However, the HSn and HSa models predict that hydrogen collisions are important, and in the latter case predict that Balmer line wings form in LTE conditions.  It should be noted that the latter two models have severe problems in their formulation; in particular, they are based on cross sections which have physically incorrect thresholds.  On the other hand, the HD model does not compare well with experiment at high energies, and the HSnc is essentially a correction to the HSn model with no other physical justification than to obtain the correct threshold.  It is therefore urgent that more physically sound theoretical results or experimental data are obtained for these processes at near threshold energies.  Excitation and, particularly, ionisation from the lowest lying states are most vital.  

Naturally, these results cast doubt on the usefulness of the Balmer line wings as  reliable temperature indicators for late-type stars.  The presumed advantage of Balmer lines has been that they are spectroscopic, and therefore unaffected by reddening (as opposed to photometry), and formed in LTE (as opposed to, e.g., Fe~I excitation equilibrium).  If the assumption of LTE is invalid, then without high quality atomic collision data, so that the departures from LTE can be accurately predicted, the temperatures derived from Balmer lines will be subject to errors.  The region of the Balmer line most sensitive to effective temperature, and that usually used and most important in fitting the lines, is the region around $2 \la \Delta \lambda \la 6$~\AA.  As a rough rule of thumb, a one per cent change in the inner wings of H$\alpha$ corresponds to 60-100~K, the exact number depending on stellar parameters (Fuhrmann~et~al.~\cite{fuhrmann93}; Barklem~et~al.~\cite{barklem02}).  For the cases considered here, we see that the possible departures from LTE (given by models HSnc or HD) in this region are of the order of a couple of per cent, thus leading to changes (raising) with respect to the temperatures determined in the LTE approximation of order 100~K.  

It is worth considering how such changes would impact comparison with observations.  Such corrections can probably not be confirmed or ruled out on the basis of observational constraints, in particular, agreement with other temperature indicators (e.g., Barklem~et~al.~\cite{barklem02}), given the errors in the analyses are of similar magnitude.  A slight weakening of the predicted solar Balmer line profiles would improve agreement with observations in terms of line strength, where a temperature of order 50~K too low is found using the LTE approximation for the case of the MARCS models and our adopted broadening theory  (see Barklem~et~al.~\cite{barklem02}).  However, Fuhrmann~et~al.~(\cite{fuhrmann93}), using different model atmospheres and broadening theory, have found temperatures in agreement with the known solar value.  This emphasises the difficulty in testing the atomic data via comparison with observations, due to the uncertainties in the model atmosphere and line broadening physics.  Even if semi-empirical atmospheres are used, the uncertainties due to assumptions such as 1D stratification and uncertainties in the line broadening physics remain.  Furthermore, it is somewhat inconsistent to compute for a semiempirical model that has been calculated using a different model atom, and this could possibly result in minor errors (see Fontenla~et~al.~\cite{fontenla06}).  Regarding line shape, since the very innermost parts of the wings are most significantly affected, the shapes of the lines would also be changed.  This would certainly lead to discrepancies with observed line shapes, though this could probably be resolved with a revision of mixing-length convection parameters, at least for theoretical models.  Any observational tests of the atomic data must be treated with care, due to the assumptions and additional uncertainties such those of a 1D static atmosphere (often modelled in LTE) and line broadening physics. While detailed and careful comparisons of the different atomic model predictions with observations, preferably using 3D models, would indeed be interesting, we believe that the preferred approach to dealing with this problem is to obtain improved theoretical and/or experimental data for the hydrogen collision processes, as discussed above.

\begin{acknowledgements}

This project arose from discussions with Martin Asplund and Bengt Gustafsson, and I am grateful for their input and encouragement.  I acknowledge the support of the Swedish Research Council. 

\end{acknowledgements}

\end{document}